\documentclass[reprint,amsmath,amssymb, aps]{revtex4-2}
\pdfoutput=1
\usepackage[utf8]{inputenc}
\usepackage{graphicx}
\usepackage{float}
\usepackage{lineno}
\usepackage{lipsum}  
\usepackage{xcolor}
\usepackage{braket}
\usepackage{physics}

\usepackage[colorlinks]{hyperref}

\usepackage[title]{appendix}

\begin{document}
\title{Preserving a qubit during adjacent measurements at a few micrometers distance}

\author{Sainath Motlakunta}
\email{smotlaku@uwaterloo.ca}
\author{Nikhil Kotibhaskar}
\author{Chung-You Shih}
\author{Anthony Vogliano}
\author{Darian Mclaren}
\author{Lewis Hahn}
\author{Jingwen Zhu}
\author{\\ Roland Habl\"utzel}
\author{Rajibul Islam}

\affiliation{ Institute for Quantum Computing and Department of Physics and Astronomy, University of Waterloo, Waterloo, Ontario N2L 3G1, Canada}
\date{\today}

\newcommand{\yb}{$^{171}\rm{Yb}^+\;$}
\newcommand{\paqm}{P_{\rm{AQM}}}
\newcommand{\D}[1]{$D_1^{(1#1)}$}
\newcommand{\adjfid}{$F_{1|2}\ $}
\newcommand{\adjfidnospace}{$F_{1|2}$}
\newcommand{\Sixj}[6]{ 
\begin{Bmatrix}
  #1 & #2 & #3 \\
  #4 & #5 & #6 
\end{Bmatrix}
}

\begin{abstract}
Protecting a quantum object against irreversible accidental measurements from its surroundings is necessary for controlled quantum operations.
This becomes especially challenging or unfeasible if one must simultaneously measure or reset a nearby object's quantum state, such as in quantum error correction.
In atomic systems - among the most established quantum information processing platforms - current attempts to preserve qubits against resonant laser-driven adjacent measurements waste valuable experimental resources such as coherence time or extra qubits and introduce additional errors.
Here, we demonstrate high-fidelity preservation of an `asset' ion qubit while a neighboring `process' qubit is reset or measured at a few microns distance.
We achieve $< 1\times 10^{-3}$ probability of accidental measurement of the asset qubit while the process qubit is reset, and $< 4\times 10^{-3}$ probability while applying a detection beam on the same neighbor for experimentally demonstrated fast detection times, at a distance of 6  \textmu m or four times the addressing Gaussian beam waist.
These low probabilities correspond to the preservation of the quantum state of the asset qubit with fidelities above 99.9\% (state reset) and 99.6\% (state measurement).
Our results are enabled by precise wavefront control of the addressing optical beams while utilizing a single ion as a quantum sensor of optical aberrations.
Our work demonstrates the feasibility of in-situ state reset and measurement operations, building towards enhancements in the speed and capabilities of quantum processors, such as in simulating measurement-driven quantum phases and realizing quantum error correction.
\end{abstract}

\maketitle



\begin{center}
\begin{figure}[t!]
    \includegraphics[width=0.5 \textwidth]{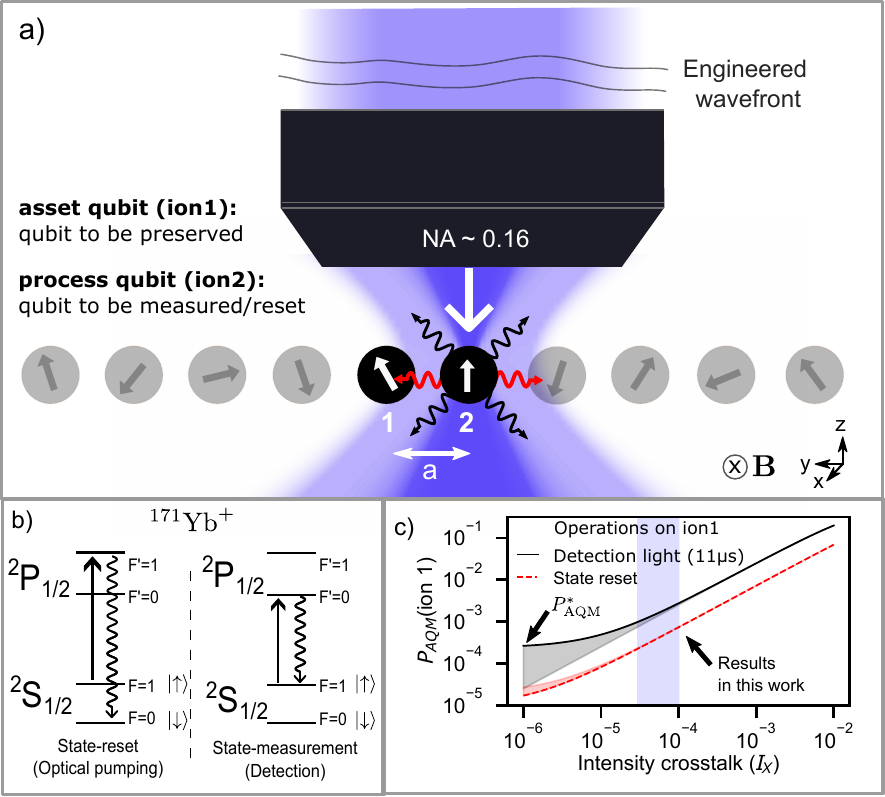} 
    \caption{\textbf{Accidental quantum measurement (AQM) of neighboring qubits.} 
    \textbf{a)} While addressing a `process' qubit (ion2) in a trapped ion chain, an `asset' qubit (ion1) at a distance $a$ away may be accidentally measured by photons that are either scattered from ion2 (red wavy lines) or from intensity crosstalk due to imperfect optical addressing (lightly shaded violet).
    AQM from imperfect optical addressing can be minimized by engineering the wavefronts incident on the microscope objective.
    \textbf{b)} Atomic transitions in \yb (Zeeman splitting not shown) for relevant incoherent processes.
    The ground state hyperfine levels $S_{1/2}\ket{F=0,m_F=0}$ and $S_{1/2}\ket{F=1,m_F=0}$ are assigned as the $\ket{\downarrow}$ and $\ket{\uparrow}$ of the effective spin-$1/2$ object or a qubit, respectively.
    Left - a quantum state is reset through optical pumping into $\ket{\downarrow}$. 
    Right - a quantum state is measured in \{$\ket{\downarrow},\ket{\uparrow}$\} basis by detecting state-dependent fluorescence\cite{Olmschenk2007ManipulationQubit} from the cycling transition.
    \textbf{c)} Calculated probability of AQM ($\paqm$) of the asset qubit (ion1) as a function of intensity crosstalk ($I_X$).
    The intensity crosstalk ($I_X=I_1/I_2$) is defined as the ratio of the optical intensity of the addressing beam on the asset qubit ($I_1$) to that on the process qubit ($I_2$).
    Here, $\paqm$ is estimated from the asset qubit's infidelity after a state detection or reset on the process qubit.
    The fidelity \cite{Nielsen2012QuantumInformation} is estimated with respect to $\ket{\uparrow}$ to represent the worst-case scenario (See supplementary information). 
    For this figure, we choose $a = 6$ \textmu m, and $I_2=I_{\mathrm{sat}}$ (the saturation intensity of the transition).
    For low crosstalk regime ($I_X<1\times10^{-5}$), inter-ion scattering sets a fundamental limit, $\paqm^*$, which can vary (shaded region) depending on the geometric properties of the system, such as the orientation of the magnetic field ($\Vec{B}$) defining the quantization axis(see Methods). 
    The results presented in this manuscript are in the regime with $I_X\lesssim  8\times10^{-5}$, leading to $\paqm < 4\times10^{-3}$ for state reset, and $\paqm<1\times10^{-3}$ with a detection beam applied for  11 \textmu s \cite{Crain2019High-speedDetectors}.
    }
    \label{fig:fig1}
\end{figure}
\end{center}

Programmable many-body quantum systems are an excellent platform for quantum information processing (QIP), including simulation of complex quantum phenomena and quantum computing.
Full programmability requires both coherent and incoherent control, such as state resets (initialization) and state measurements at the level of its individual building blocks\cite{Ladd2010QuantumComputers, DiVincenzo2000TheComputation}.
Coherent dynamics are, in principle, reversible, while incoherent operations generally constitute irreversible quantum measurements.
The ability to perform measurements and resets on a subsystem in the middle of coherent dynamics  (`mid-circuit measurements and resets') is a powerful tool for simulating new classes of quantum phenomena such as measurement-driven quantum phase transitions \cite{Noel2022Measurement-inducedComputer, Czischek2021SimulatingCircuits, Li2018QuantumTransition, Skinner2019Measurement-InducedEntanglement, Chan2019Unitary-projectiveDynamics, Sang2021Measurement-protectedPhases, Lavasani2021Measurement-inducedCircuits} and executing quantum error correction protocols  \cite{Ladd2010QuantumComputers, Ryan-Anderson2021RealizationCorrection, Shor1995SchemeMemory}.
A primary challenge  \cite{Gaebler2021SuppressionMicromotion} of subsystem mid-circuit measurement and reset is the accidental quantum measurement (AQM) of the remaining system during the process, leading to irreparable decohering errors. 
In particular, for atomic quantum systems such as trapped ions, state reset, and measurement are performed via resonant laser beam illumination, and accidental scattering of photons leads to a finite probability of AQM ($\paqm$).
This probability can be prohibitively high, as the typical inter-atomic separation is comparable to the optical resolution.
To mitigate the high probability of AQM in atomic QIP experiments, strategies such as physical separation of atoms through shuttling  \cite{Crain2019High-speedDetectors, Pino2021DemonstrationArchitecture, Zhu2021InteractiveAdvantage, Wan2019QuantumProcessor}, usage of additional ancilla qubits of the same  \cite{Noel2022Measurement-inducedComputer} or different atomic species  \cite{Negnevitsky2018RepeatedRegister, Bruzewicz2017High-FidelityChain, Home2009CompleteProcessing, Singh2023Mid-circuitQubits}, hiding \cite{Riebe2004DeterministicAtoms, Manovitz2022Trapped-IonFeedback, Hilder2022Fault-TolerantComputer, Schindler2013AIons, Lis2023Mid-circuitArrays} qubits to states outside the computational Hilbert space or adopting other suppression techniques \cite{Gaebler2021SuppressionMicromotion} are employed. 
However, these techniques waste resources (circuit time, extra qubits) and introduce additional errors (e.g., errors due to motional heating or imperfect coherent operations).

Here, we demonstrate preservation of the quantum state of an `asset' ion qubit with high fidelity while a neighboring `process' qubit at a few microns distance is reset or measured.
We achieve $\paqm < 1\times 10^{-3}$ of the asset qubit during the process qubit reset operation and $\paqm < 4\times 10^{-3}$ while applying detection light on the process qubit for experimentally demonstrated \cite{Crain2019High-speedDetectors} fast detection times.
These low probabilities of AQM correspond to retaining the quantum state of the asset qubit to above 99.9\% and 99.6\% fidelities for the reset and measurement processes, respectively.
Our explorations, presented here, further provide a framework to optimize multidimensional optical parameters for maximizing in-situ operation fidelity.
Our in-situ incoherent operations are enabled by exquisite control over optical wavefronts from a holographic addressing system  \cite{Shih2021ReprogrammableControl} that compensates for aberrations sensed using a single ion's quantum state.
This approach relies on robust optical engineering rather than special trapping architectures and can be adapted to other atomic QIP systems.
Our demonstrated high-fidelity results immediately enable explorations of novel measurement-driven quantum simulation protocols  \cite{Czischek2021SimulatingCircuits} and open quantum systems, such as quantum simulations with local dissipation and measurements, and quantum reservoir engineering.
The in-situ operations lead to scalable, simple, robust, and fast QIP protocols compared to other error mitigation techniques like shuttling and usage of ancilla for mid-circuit measurements.

Figure \ref{fig:fig1}a describes the two mechanisms for AQM of the asset qubit (ion1) during a measurement or reset on the process qubit(ion2): finite intensity crosstalk (due to imperfect optical addressing) and inter-ion scattering (absorption of photons emitted by ion2).
Here, we ignore any additional measurement arising due to the entanglement of qubits in the system.
While both the state reset and detection (Fig. \ref{fig:fig1}b) generally constitute quantum measurements, in practice, detection involves at least an order of magnitude more scattered photons due to the inefficiency of the measurement apparatus.
Due to the increased number of scattered photons as well as finite intensity crosstalk from the laser beam, $\paqm$ will increase with detection time.
However, advances in state measurement techniques and protocols have enabled high-fidelity detection in $\sim 10$ \textmu s  \cite{Crain2019High-speedDetectors}. 
The fast detection times allow in-situ, site-selective measurement of the process qubit with $\paqm \sim 10^{-3}$ for the asset qubit as long as the intensity crosstalk can be contained below $10^{-4}$ level (Fig. \ref{fig:fig1}c).
Inter-ion scattering sets a fundamental limit  to $\paqm$ of $\paqm^*\sim 1/a^2$ in the regime where inter-ion spacing $a$ is much larger than the wavelength of radiation.
In addition, the exact value of $\paqm^*$ will depend on the magnetic field (quantization axis) configuration (See methods).
We find, by atomic physics calculations as well as experiments, that the $\paqm$ with our measured intensity crosstalk of $I_X\lesssim 8\times 10^{-5}$ at the asset qubit approaches but is not yet limited by $\paqm^*$.

We use the ground state hyperfine levels of \yb ions trapped and Doppler-cooled in a `four-rod' Paul trap as $\ket{\downarrow}$ and $\ket{\uparrow}$ of the effective spin-$1/2$ object or a qubit.
These ions are individually probed through an addressing system with an effective numerical aperture of 0.16(1).
The optical aberrations in the system are characterized (see Methods) using a single ion as a quantum sensor.
Using a measured aberration phase profile, a Fourier hologram employed on a digital micromirror device (DMD) is programmed to create a diffraction-limited Gaussian beam of waist $w$=1.50(5) \textmu m in the ion plane.
This beam is positioned at a programmable distance $d$ from the ion while minimizing intensity leakage onto neighboring ions.

\begin{center}
\begin{figure}[t!]
    \includegraphics[width =0.5 \textwidth]{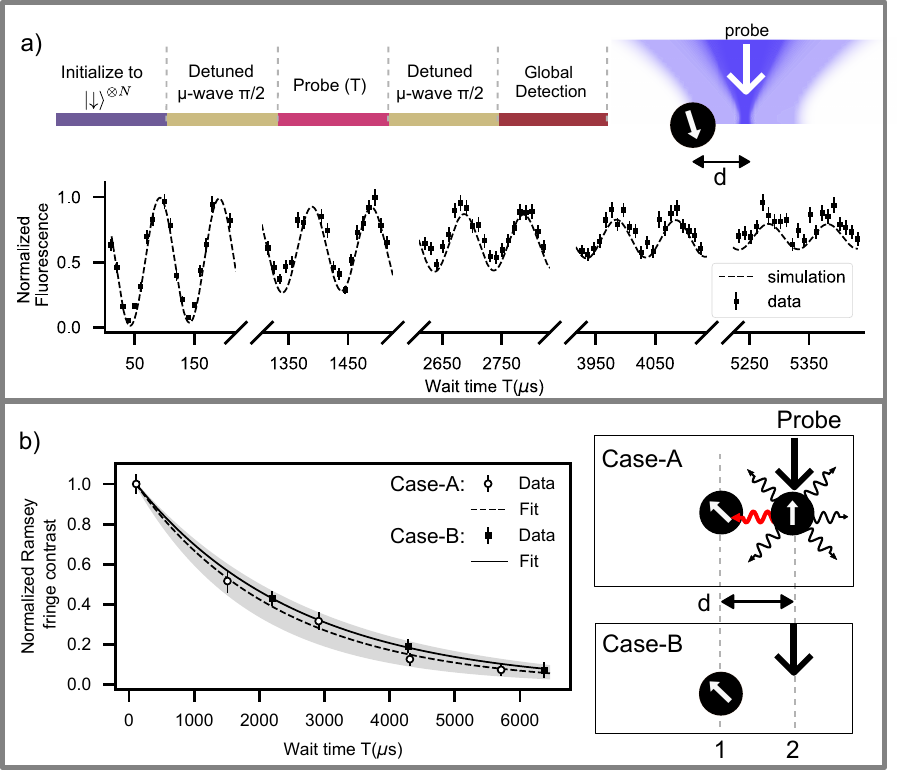}
    \caption{\textbf{AQM characterization scheme.} 
    \textbf{a)} (Top) Ramsey interferometric protocol (see Supplementary information) to measure qubit coherence time $T_2^*$ when the incoherent probe beam is applied for time $T$ at a distance $d$.
    (Bottom) Ramsey fringes in normalized fluorescence originating from the detuning between the microwave source and the qubit frequency. 
    $T_2^*$ is extracted from the decay in Ramsey fringe contrast.
    Here, the background-subtracted fluorescence counts are measured during the global detection step and are normalized with respect to the counts from $\ket{\uparrow}$.
    The data shown here are for a single ion $(N=1)$ illuminated with the state-detection probe light ($I=1.25(16)I_{\mathrm{sat}}$, waist $w$=1.50(5) \textmu m) at a distance of $d=6.0(3)w=9.0(4)$ \textmu m and Ramsey detuning of 10 kHz.
    Error bars indicate standard error from 200 experimental repetitions. 
    Intensity crosstalk $I_X$ is estimated from numerical simulations of the master equation from the measured $T_2^*$. We find, using numerical simulations solving the master equation of the system (dashed line, see supplementary information), that the intensity cross-talk, $I_X= 3.4(6)\times 10^{-5}$ for this data.
    \textbf{b)} Comparison of Ramsey fringe decay profiles between case-A: an ion located at the probe beam focus $(N=2)$, and case-B: no ion at probe beam focus $(N=1)$.
    Data points represent Ramsey fringe contrast measured over two fringes, and the fits are exponential decay with $T_2^*$ as a fitting parameter.
    The Ramsey fringe contrast is normalized with the contrast measured at $T\approx 0$.
    Error bars denote standard deviation in estimating Ramsey fringe contrast, using 20 bootstrapping repetitions from 200 measurements.
    The shaded region indicates fluctuations of experimental settings over periodic calibration of the probe beam location with respect to ion2  (see supplementary information) for case-A. 
    $T_2^*$ values measured for case-A and case-B lie within the error bounds, indicating that the decoherence is limited by the intensity crosstalk and not by inter-ion scattering.
     }
    \label{fig:fig2}
\end{figure}
\end{center}

In the regime where the probability of the asset qubit accidentally scattering a photon, $\paqm\ll 1$, we find numerically that the infidelity of the asset qubit is a good estimate of $\paqm$ (Fig. \ref{fig:fig1}c). 
The fidelity \cite{Nielsen2012QuantumInformation} of preserving the state of asset qubit is estimated from fringe contrast in a Ramsey interferometry experiment (Fig.  \ref{fig:fig2}a.)
We  measure the fringe decay (decoherence) time $T^*_2$ of the asset qubit(ion1) and estimate (see Supplementary information) the fidelity of preserving its state after measurement or reset on the process qubit (ion2) from,
\begin{equation}
    \label{eqn:eqn1}
    F_{1|2} = \frac{2}{3}\exp\left[-\frac{\tau(\rm{ion2})}{T^*_2(\rm{ion1})}\right]+\frac{1}{3}. 
\end{equation} 
Here, $\tau$(ion2) is the time for which the resonant probe beam illuminates the process qubit. 
From the measured $T^*_2$, we estimate the intensity of probe light sampled by the asset qubit.
The long quantum memory of the asset qubit ($T^*_2$ without any probe light is at least two orders of magnitude longer than the results in Fig. \ref{fig:fig2}b) enables it to act as a sensitive, high-dynamic range sensor for intensity crosstalk.

%
To distinguish the decoherence caused by inter-ion scattering and the imperfect optical addressing, we perform the above Ramsey measurements for two different cases.
case-A uses two ions, separated by a distance $d$, with a probe beam parked on ion2.
case-B uses only one ion with a probe beam parked at the same distance $d$ from the ion (Fig.  \ref{fig:fig2}b).
For $d = 6w$ (9 \textmu m), we find that the Ramsey fringe decay time ($T_2^*$) for both experiments is indistinguishable (within the experimental fluctuations)(Fig.  \ref{fig:fig2}b).
This verifies that the inter-ion scattering is not the major source of decoherence in our experiment.
Thus we could use a single-ion (case-B) to quantify the fidelity \adjfid in our addressing scheme, which greatly simplifies the measurement scheme.

The process duration $\tau$ and the decoherence time $T_2^*$ in Eq.\ref{eqn:eqn1} may have different dependence or optima over optical parameters (such as polarizations, spectral purity, etc.).
In the following experiments, we maximize \adjfid by maximizing the fraction of light contributing to the process (state reset or measurement) while minimizing (where possible) the fraction of light that accidentally measures the asset qubit.
\begin{figure*}[ht!]
    \includegraphics[width =1 \textwidth]{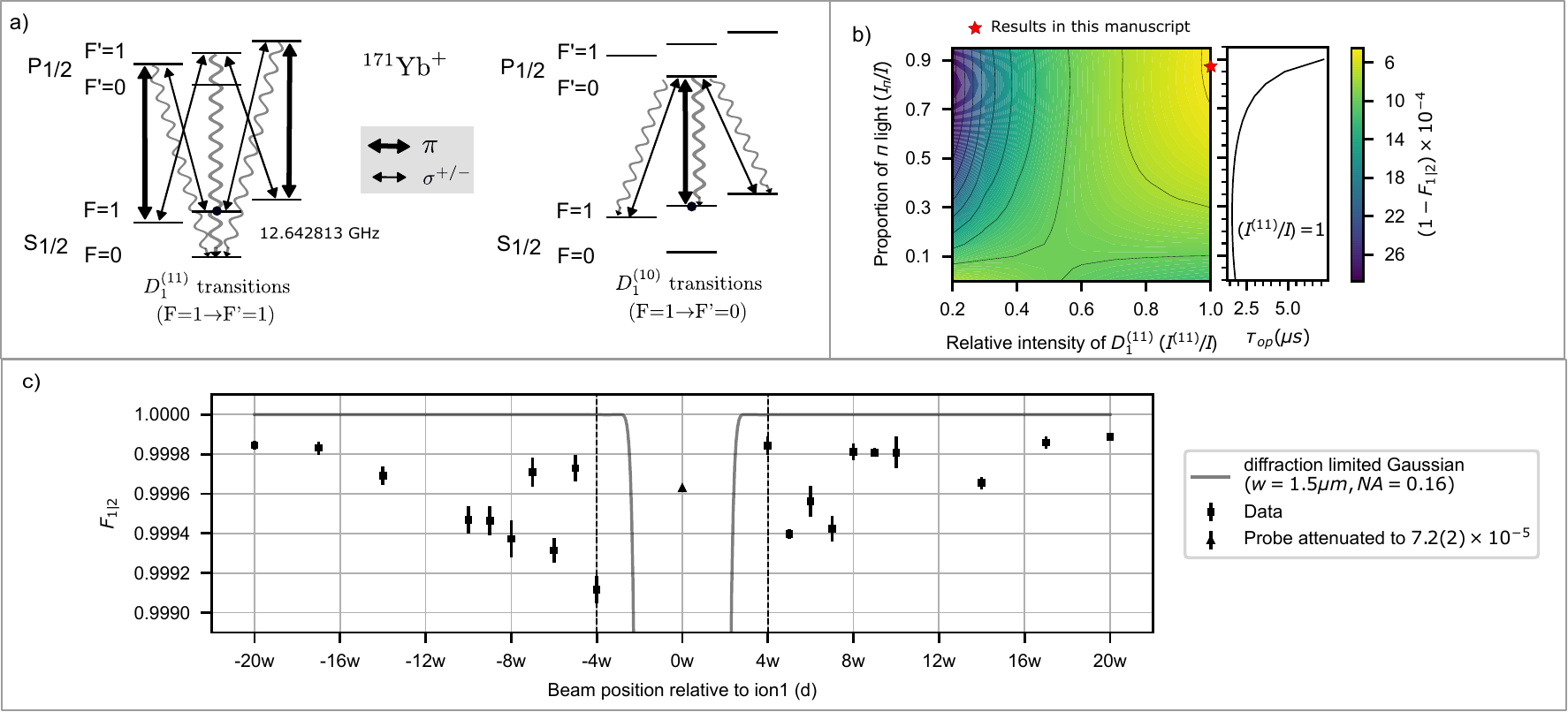}
    \caption{\textbf{Fidelity (\adjfidnospace) of preserving ion1 for state reset light at ion2 location.} 
    \textbf{a)} Excitation and decay mechanisms for the \D1 and \D0 transitions in \yb ion initialized in state $\ket{\uparrow}$ \cite{Olmschenk2007ManipulationQubit}, for various polarizations (thick arrows representing $\pi$ and thin arrows representing $\sigma^{\pm}$).
    The \D1 transitions contribute to state reset via optical pumping, although any residual \D0 light (e.g., from frequency modulation via an electro-optic modulator\cite{Olmschenk2007ManipulationQubit}) may degrade \adjfid.
    \textbf{b)} Calculated \adjfid for the state reset process as a function of the ratio of the intensity of \D1 component $I^{(11)}$ to the total intensity $I$ (where $I=I^{(11)}+I^{(10)}$ with $I^{(10)}$ indicating \D0 component)
    and ratio of the intensity of $\pi$ polarization $I_{\pi}$ to the total intensity $I$ (where $I=I_{\pi}+I_{\sigma^+}+I_{\sigma^-}$ with equal intensities in $\sigma^+$ and $\sigma^-$ polarizations)
    Here, \adjfid is calculated using numerical simulations of the master equation (see supplementary information) under the conditions of $I_2 = 1.25I_{\rm{sat}}$ and $I_X = 5\times10^{-5}$.
    The red star marker indicates the parameters used to measure \adjfid in (c).
    Additionally, the plot on the right shows an estimation of state reset times $\tau_{\mathrm{op}}(\rm{ion2})$ as a function of $I_{\pi}/I$ for $I^{(11)}/I=1$
    \textbf{c)} \adjfid vs  $d$ expressed in multiples of the beam waist $w$ (case-B in Fig. \ref{fig:fig2}b).
    Here, $w = 1.50(5)$ \textmu m is the Gaussian beam waist for the addressing beam.
    Error bars denote standard deviation in estimating \adjfid, using 20 bootstrapping repetitions from 200 measurements (See supplementary information).
    For calibrating crosstalk $I_X$, we measure \adjfid for a probe beam with relative intensity attenuated to $7.2(2)\times10^{-5}$ addressing ion1 (triangle marker at $d=0$).
    For comparison, \adjfid is calculated (solid gray line) for a diffraction-limited (NA = 0.16) Gaussian beam of beam waist $w = 1.50$ \textmu m.
    \adjfid is $>$99.9$\%$ for $d\ge 4w$ (see discussion).
    For these measurements, $I_2 = 1.25(16)I_{\mathrm{sat}}$, $I_{\pi}/I = 0.86$, $I^{(11)}/I = 1$, $\tau_{\rm{op}}=9.73(7)$ \textmu s. 
    }
    \label{fig:fig3}
\end{figure*}

\begin{figure*}[ht!]
    \includegraphics[width =1 \textwidth]{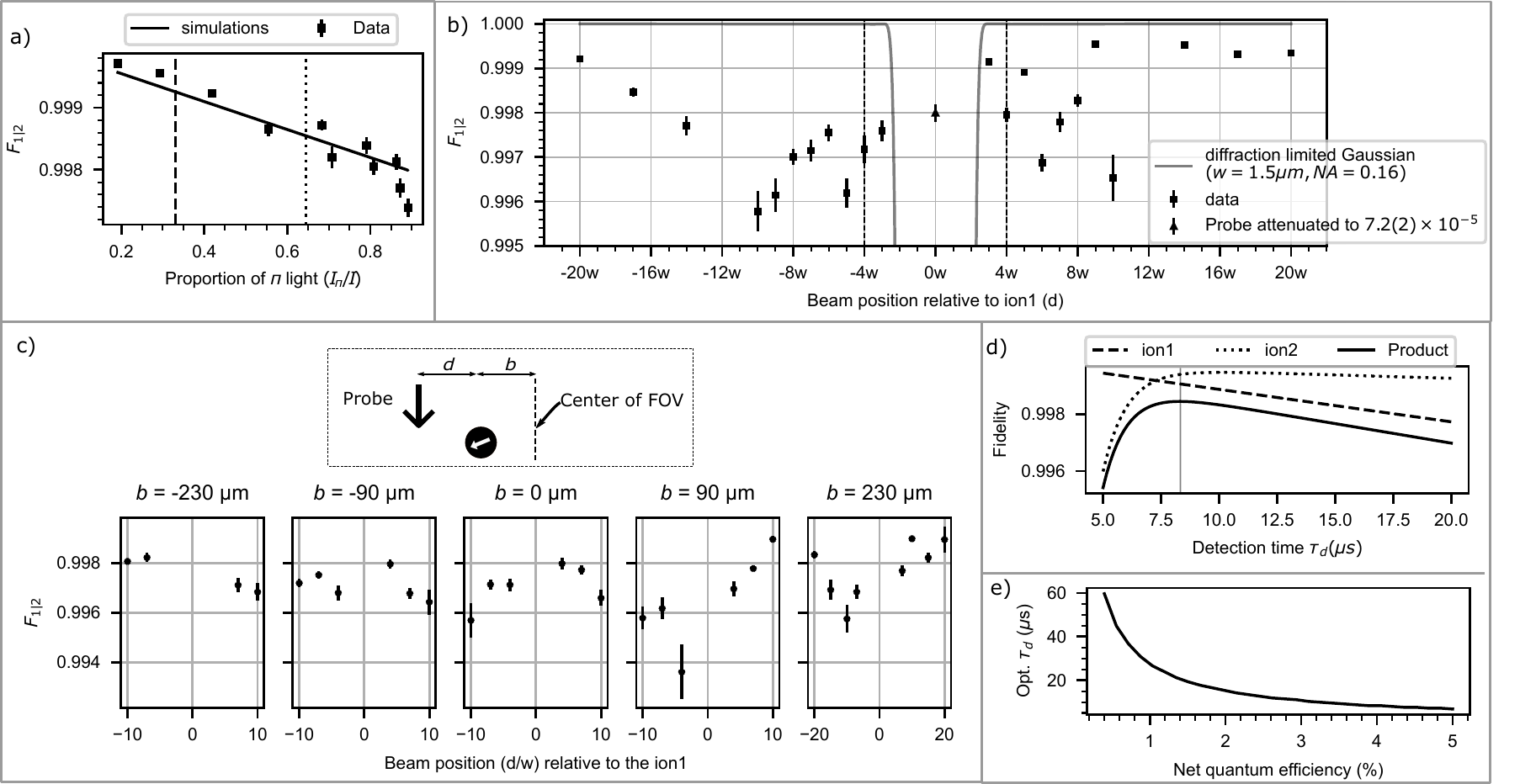}
    \caption{\textbf{Fidelity (\adjfidnospace) of preserving ion1 for detection light at ion2 location.} 
    \textbf{a)} \adjfid vs polarization of the detection probe light, showing that it is maximized for probe light with no $\pi-$polarization.
    The dashed line represents the optimal polarization  \cite{Ejtemaee2010OptimizationDetection} for the process qubit (ion2) state-detection.
    The dotted line represents the polarization used to measure \adjfid in parts \textit{b-c}.
    Measured values of \adjfid at $d=4w$, shown in parts \textit{a-c}, are for detection probe light of intensity $I=1.25(16)I_{\mathrm{sat}}$ applied for $\tau_{\mathrm{d}}=$ 11 \textmu s.
    Error bars in parts \textit{a-c} denote standard deviation in estimating \adjfid, using 20 bootstrapping repetitions from 200 measurements (See supplementary information).
    \textbf{b)}\adjfid vs the distance $d$ (case-B in Fig. \ref{fig:fig2}b).
    For comparing the crosstalk $I_X$, we measure \adjfid for a probe beam with relative intensity attenuated to $7.2(2)\times10^{-5}$ addressing ion1 (triangle marker at $d=0$).
    For comparison, \adjfid is calculated(solid gray line) for a diffraction-limited (NA=0.16) Gaussian beam of beam waist $w$,
    \adjfid fidelity is $>$99.6$\%$ for $d\ge 4w$.
    \textbf{c)} Measured \adjfid for various shifted locations of the ion from the center of the field of view (FOV). Here the center of FOV denotes the location at which the aberrations have been characterized and compensated (see methods).
    \adjfid is preserved for a large FOV of 460 \textmu m.
    \textbf{d)} Calculated process qubit (ion2) detection fidelity\cite{MarkActon2008DetectionAtoms,Crain2019High-speedDetectors} (See supplementary information) and asset qubit(ion1) preservation fidelity (\adjfid) as the function of $\tau_{\mathrm{d}}$ (detection time).
    Here, for estimating the process qubit (ion2) detection fidelity, we assume that the process qubit is illuminated with a detection beam of $I_2=I_{\mathrm{sat}}$ with optimal polarization and a measurement apparatus of net detection efficiency of 4\%, compatible with the state-of-the-art experiments.
    We employ a photon count thresholding method to differentiate between $\ket{\uparrow}$ and $\ket{\downarrow}$ states. Furthermore, we use an algorithm that completes the detection process upon measuring the first photon, reducing detection time by a factor of 2  \cite{Crain2019High-speedDetectors, Hume2007High-FidelityMeasurements, Langer2006HighIons}.
    For estimating \adjfid, we assume intensity crosstalk of $I_X=5\times 10^{-5}$, $I_2=I_{\mathrm{sat}}$, and optimal polarization for the process qubit state-detection.
    The vertical line at $\tau_{\mathrm{d}}\approx8.5$ \textmu s represents the optimal detection time that maximizes the product of these two fidelities.
    \textbf{e)} Optimal detection time (opt. $\tau_{\mathrm{d}}$) as a function of the net detection efficiency of the measurement apparatus.
    }
    \label{fig:fig4}
\end{figure*}

\textbf{Site-selective state reset - }
The process of state reset through optical pumping is done by using a probe that drives the transition \D1 from $S_{1/2} \ket{F=1}$ to $P_{1/2}\ket{F'=1}$ \cite{Olmschenk2007ManipulationQubit} (Fig.  \ref{fig:fig3}a).
We choose the process time $\tau_{\mathrm{op}}$(ion2) $=  7T_1$, where $T_1$ indicates the time at which the normalized ion fluorescence of ion2 drops to $1/e$ compared to its initial value.
This will ideally reset the quantum state of the process qubit to $\ket{\downarrow}$ with the fidelity of $1-e^{-7} = 0.999$.
Note that spectral components of light apart from \D1 may decohere the asset qubit while not contributing to the reset on the process qubit (ion2).
For example, optical pumping light derived by frequency modulation (electro-optic modulation) employed in typical ion trap experiments  \cite{Olmschenk2007ManipulationQubit} contains residual \D0 component (the spectral component used for detection).
This \D0 component will increase $\paqm$ and hence reduce \adjfid, as shown by numerical simulation data in Fig. \ref{fig:fig3}b where \adjfid is maximized for a probe with relatively higher intensity in \D1 component ($I^{(11)}$).

Probe light of different polarizations has unequal contributions to the asset qubit's fidelity \adjfid for the case of state reset. 
For \D1 transition, since the $\ket{\uparrow}=S_{1/2}\ket{F=1,m_F=0}$ to $P_{1/2}\ket{F'=1,m_F'=0}$ is dipole forbidden, the component with $\pi$ polarization with intensity $I_\pi$ doesn't contribute to the AQM of the asset qubit. 
So, for the case of probe with $I^{(11)}/I = 1$, \adjfid increases with $I_\pi/I$ (Fig. \ref{fig:fig3}b). 
This increase in \adjfid comes at the cost of increasing the state reset $\tau_{\mathrm{op}}$ of ion2. 
For the case of $I^{(11)}/I \neq 1$, the $\pi$ polarizations of the \D0 transition component of the probe still contribute to the AQM of the asset qubit.
Hence an increase in $I_\pi/I$ decreases \adjfid as the light with $\pi$ polarization only contributes to AQMs of the asset qubit but not to the reset of the process qubit(Fig.  \ref{fig:fig3}b).

With the polarization of the state reset beam optimized, we characterize \adjfid (in case-B configuration) as a function of beam position relative to the ion ($d$) (Fig.   \ref{fig:fig3}c) and observe $F_{2|1}>99.90\%$ for $d\ge 4w$.
To calibrate the intensity crosstalk for these measurements, we perform another experiment with $d=0$ with attenuated intensity.
\adjfid measured with attenuated light confirms that our intensity crosstalk is in the regime of $\lesssim8\times10^{-5}$ (Fig. \ref{fig:fig3}c).
This estimation of the intensity of crosstalk is also corroborated by our atomic physics simulations (see supplementary information).

\textbf{Site-selective state measurement - }
State measurement of the ion qubits is achieved by detection of the state-dependent fluorescence  \cite{Olmschenk2007ManipulationQubit}.
The ions are excited by light resonant to the \D0 cycling transition and the light they scatter is detected with finite efficiency by a measurement apparatus.
The duration of this measurement (detection time $\tau_{\mathrm{d}}$) is chosen such that the state of the ion can be inferred with high fidelity.
Detection time is highly dependent on the efficiency of the measurement apparatus, and a detection time of as low as 11  \textmu s with state-detection fidelities as high as 99.931(6)\% has been recently demonstrated for \yb trapped ions  \cite{Crain2019High-speedDetectors}.
Hence we report \adjfid when a detection beam was applied on ion2 location for a time of 11 \textmu s (Fig.  \ref{fig:fig4}).
Note that the detection beam parameters used in Ref. \cite{Crain2019High-speedDetectors} are compatible with our detection beam.

$\sigma^+$ and $\sigma^-$-polarized \D0 light don't cause AQMs to the asset qubit in $S_{1/2}\ket{F=1,m_F=0}$ state, ignoring low probability off-resonant excitation (Fig.  \ref{fig:fig3}a).
Hence, the asset qubit's fidelity is maximized with the least proportion of $\pi$ light (Fig.  \ref{fig:fig4}a). 
But the optimal polarization \cite{Ejtemaee2010OptimizationDetection} for the highest scattering rate and hence the highest detection fidelity for ion2 is $I_\pi = I_{(\sigma^+)} = I_{(\sigma^-)}$.

We examine \adjfid as a function of beam position relative to the ion ($d$) (Fig. \ref{fig:fig4}b) using a detection beam in a case-B configuration.
We find that the long coherence times ($T_2^*$) result in fidelities \adjfid$>99.5\%$ for $d\ge4w$ and \adjfid$>99.9\%$ for $d\ge20w$. 
Note that the polarization of the probe beam for these measurements is $I_\pi\approx0.6$ and even higher fidelities could be acheived for the optimal detection polarization (Fig.  \ref{fig:fig4}a).
This high fidelity is maintained in the measurements with ion shifted ~100  \textmu m and ~200  \textmu m away from the center of the field of view (FOV) (Fig.  \ref{fig:fig4}c), demonstrating that in-situ measurements are possible in a long chain of ions.
Note that, for all the aforementioned measurements, the aberration was compensated using the phase profile measured at $b$=0 (See Methods)
Moreover, by compensating the aberrations using a phase profile measured at a different point located away from the center of the field of view, it is possible to achieve even higher fidelities at that specific point.

The detection-fidelity of the process qubit for a given detection efficiency (from a limited NA and photon collection loss) increases with increasing the detection time ($\tau_{\mathrm{d}}$) (ignoring the off-resonant effects) whereas \adjfid decreases (Fig. \ref{fig:fig4}d).
The optimal detection time depends on the relative importance of these fidelities in a given quantum algorithm.
For example, one metric to find optimal detection time could be to maximize the product of these fidelities.
This optimal detection time is highly dependent on the net efficiency of the detection apparatus (Fig. \ref{fig:fig4}e).





\textbf{Discussion - }
In summary, we have demonstrated high fidelity in preserving an ion qubit while the neighboring qubit is reset or measured at a few microns distance.
Our results are comparable to the state-of-the-art QIP experiments  \cite{Chertkov2022HolographicComputer, Gaebler2021SuppressionMicromotion, Crain2019High-speedDetectors} that employ shuttling of qubits to be preserved away from reset or measurement laser beams by hundreds of microns distance.

Further, our protocol could be combined with other error-mitigation methods, such as shorter-distance shuttling or usage of a different isotope of the same ion species, paving ways to reduce crosstalk errors compatible with quantum error correction protocols.
Short-distance (tens of microns) shuttling would also improve the speed of the quantum algorithms and reduce errors from motional heating when compared to hundreds of microns shuttling used in current experiments.
For a typical isotope shift of a few hundred MHz and our demonstrated $\lesssim 8\times 10^{-5}$ intensity crosstalk, the $\paqm$ for state reset and measurement can be reduced to the $10^{-6}$ level.
Using a different isotope of the same ion species will also remove challenges, such as reduced motional coupling between ions of disparate masses \cite{Teoh2021ManipulatingTweezers, Inlek2017MultispeciesNetworking, Sosnova2021CharacterChains} during mid-circuit sympathetic cooling and quantum gate operations.

Our crosstalk measurement scheme employs temporal separation of probe light illumination and detection of an ion qubit and hence overcomes sensitivity limitations due to unwanted background scattering of resonant light from optics leaking onto photon detectors in previous experiments \cite{Shih2021ReprogrammableControl}.
This in turn allows measurement of crosstalk over a large dynamic range.

Ions are localized to $<$ 100 nm at typical laser-cooling temperatures and trap frequencies, making it possible to characterize aberrations with the ion sensor for larger numerical aperture (NA) systems.
With large NA, the beam waist $w$ decreases, thus the ion separations can be decreased without increasing $\paqm$ to achieve higher qubit-qubit interaction strengths \cite{Monroe2021ProgrammableIons}.

Our demonstrated high fidelity over a field of view (FOV) of 450  \textmu m corresponds to $\sim$ 50 ions in a linear chain for typical harmonic trapping parameters (radial trap frequency of approx. $2\pi \times  5 \;\rm {MHz}$ and axial trap frequency of approx. $2\pi \times 30\;\rm{kHz}$).
The slight decay of fidelity away from the center of FOV can be compensated by recalibrating aberrations away from the center.
However, even without extra calibrations, the fidelity \adjfid can be maintained over the entire chain, as inter-ion separation away from the center of an ion chain also increases in a harmonic trap (from $4w=6$ \textmu m at the center becoming $\approx 10w=15$ \textmu m near the edge for parameters above \cite{Steane1997TheProcessor}). 

For typical radiofrequency ion traps (e.g., surface traps \cite{Maunz2016High2.0.}, `blade' electrode traps\cite{He2021AnDirections}), NA $>$ 0.5 is accessible for photon collection simultaneously with NA $\sim$ 0.3 (in a perpendicular direction) for optical addressing, allowing for independent optimization for photon collection and addressing.
While high quantum efficiency and negligible dark counts make $\sim 10$ \textmu s detection time possible \cite{Crain2019High-speedDetectors}, less-expensive photomultiplier tubes (PMT) can also allow $\sim$ 20 \textmu s detection time (Fig.  \ref{fig:fig4}e)  \cite{MarkActon2008DetectionAtoms, Crain2019High-speedDetectors} under otherwise identical conditions for maintaining high asset qubit preservation fidelities of $>99.2\%$. 

While the asset qubit coherence in our measurements is limited by intensity crosstalk, the $\paqm^*$ from inter-ion scattering for state detection may be suppressed even further with the proper choice of the local magnetic field.
For \yb, it is possible to suppress (see methods) the intensity of $\pi$ light scattered from the process qubit in the direction of the asset qubit by aligning the magnetic field along the ion chain \cite{Budker2004AtomicSolutions}, thereby maximizing \adjfid (Fig.  \ref{fig:fig4}a).
In contrast, the optimal orientation of the magnetic field for state reset is perpendicular to the ion chain.

Comparing with the inter-ion scattering calculations in Ref.  \cite{Bruzewicz2017High-FidelityChain}, we find that the insensitivity to $\sigma^{\pm}$ photons (for state-detection through \D0 transition) for \yb gives about $\sim 2$ times reduction in $\paqm$ compared to some other species, such as $^{40}\rm{Ca}^+\;$ that is affected by all polarizations.
Our scheme of obtaining low $\paqm$ can be easily adapted to other ion species or different QIP platforms that benefit from high-quality individual optical addressing.

\section*{Author contributions}
SM, C-YS, AV, LH, and JZ performed the experiments following an initial feasibility study by RH, C-YS, and SM.
SM, NK, and DM performed theory calculations, numerical simulations, and analyses.
SM, AV, and RI wrote the manuscript with inputs from all authors.
All the authors contributed to the scientific discussions.
RI supervised the whole project.

\section*{Acknowledgements}
We thank Yu-Ting Chen and  Crystal Senko for scientific discussions.
We acknowledge financial support from the Canada First Research Excellence Fund (CFREF), the Natural Sciences and Engineering Research Council of Canada (NSERC) Discovery program (RGPIN-2018-05250), the Government of Canada’s New
Frontiers in Research Fund (NFRF), Ontario Early Researcher Award, University of Waterloo, and Innovation, Science and Economic Development Canada (ISED).
\clearpage
\section{Methods}
\textbf{$\paqm^*$ due to inter-ion scattering - }
\label{sec:inter-ion_scattering_calc}
Consider two ions (ion1 and ion2) separated by a distance $a$ in an ion chain. 
An ideal resonant laser beam illuminates ion2 without leaking any photons onto ion1. 
ion2 scatters photons at a rate $\Gamma_{\mathrm{sc}}(\mathrm{ion2})$, a portion of which are incident on ion1. 
The effective intensity of light on ion1 from these scattered photons is denoted by $I_{\mathrm{ab}}(\mathrm{ion1})$.
The relation between $\Gamma_{\mathrm{sc}}(\mathrm{ion2})$ and $I_{\mathrm{ab}}(\mathrm{ion1})$ is given as  
$$ I_{\mathrm{ab}}(\mathrm{ion1}) = f_{\mathrm{pol}} f_{\mathrm{angle}}\frac{h\nu\Gamma_{\mathrm{sc}}(\mathrm{ion2})}{4\pi a^2}$$
Here, $\nu$ represents the frequency of the scattered light, $f_{\mathrm{pol}}$ denotes the fraction of light whose polarization affects ion1, and $f_{\mathrm{angle}}$ represents the angular dependence of the scattered light.

When ion2 is illuminated with an ideal state-detection beam, it emits light of all polarizations. However, only the $\pi$ polarization causes $\paqm^*$ in ion1. Therefore, we have $f_{\mathrm{pol}} = 1/3$.
The angular dependence of light scattered in $\pi$ polarization by ion2 in the direction of ion1 is given by $f_{\mathrm{angle}} = \cos^2(\theta)$ \cite{Budker2004AtomicSolutions}, where $\theta$ is the angle between the magnetic field and the ion chain.
In our setup, since the magnetic field is perpendicular to the ion chain, $f_{\mathrm{angle}}=1$.
However, by choosing the magnetic field along the ion chain, $f_{\mathrm{angle}}$ can be suppressed to zero.
For a state-detection probe beam with an intensity of $I_2 = I_{\mathrm{sat}}$, we use the optimal scattering rate of ion2 \cite{Noek2013HighQubit} to estimate $\Gamma_{\mathrm{sc}}(\mathrm{ion2})$.
Assuming an inter ion spacing of 6  \textmu m, we estimate that $I_{\mathrm{ab}}(\mathrm{ion1})\approx 9.5\times10^{-6}I_{\mathrm{sat}}$.
This results in a $\paqm^* = 2\times 10^{-4}$ for $11$ \textmu s state-detection.

For a state reset operation on ion2, $f_{\mathrm{pol}}=2/3$ since light with both $\sigma^+$and $\sigma^-$ polarizations affect ion1.
Additionally, for the case of a magnetic field perpendicular to the ion chain, we have $f_{\mathrm{angle}}=1/2$ \cite{Budker2004AtomicSolutions}.
We estimate that $I_{\mathrm{ab}}(\mathrm{ion1})\approx 1.3\times10^{-6}I_{\mathrm{sat}}$ for the state reset operation.
This results in a $\paqm^* = 1\times 10^{-5}$ for state reset.

\textbf{Aberration correction - }
\label{sec:aberration_correction}
We characterize optical aberrations in the entire beam path in terms of a Fourier plane (FP) phase map.
An amplitude hologram on a Digital Micromirror Device (DMD) in the FP allows us to control the amplitude and phase of the diffracted light.
The relative optical phase between two FP `patches' is measured from the interference of beams that are diffracted from these patches.
We use a single ion as a quantum sensor to measure this interference signal.
We use an optical pumping beam on the ion, initialized in state $\ket{\uparrow}$, and observe state-dependent fluorescence signal from the ion as it gets pumped into state $\ket{\downarrow}$.
By varying the phase of one of the FP patches and observing ion fluorescence for a fixed optical pumping time, we extract the interference profile and hence the relative phase.
This approach is highly sensitive, as only a few photons are needed for optical pumping, allowing us to map out the phase diagram for the entire FP with very low optical power ($\sim 200$ \textmu W of 369 nm light).
By decoupling the probing and measurement, we achieve a higher signal-to-noise ratio compared with our previous approach  \cite{Shih2021ReprogrammableControl}, where an unwanted scattering of the probe beam from optics leaking onto the detector was a limiting factor.
The aberrations are then compensated by generating the corrective hologram on the DMD using an iterative Fourier transform algorithm (IFTA) \cite{Shih2021ReprogrammableControl}.

\textbf{Intensity and polarization calibration - }
We collect the ion fluorescence time-series data from many optical pumping experiments, where we controllably vary the relative optical power and polarization of the optical pumping light between experiments.
Trends in these time-series data are fitted by numerical simulation to extract the saturation parameter and polarization of the light illuminating the ion (See Supplementary information).

\newpage
\section*{Supplementary information}
\begin{appendices}
\renewcommand\thefigure{S\arabic{figure}} 
\setcounter{figure}{0}    
\renewcommand\theequation{S\arabic{equation}} 
\setcounter{equation}{0}
\renewcommand\thesubsection{\arabic{subsection}} 
\setcounter{subsection}{0} 

\begin{figure*}[t!]
    \centering
    \includegraphics[width =0.9 \textwidth]{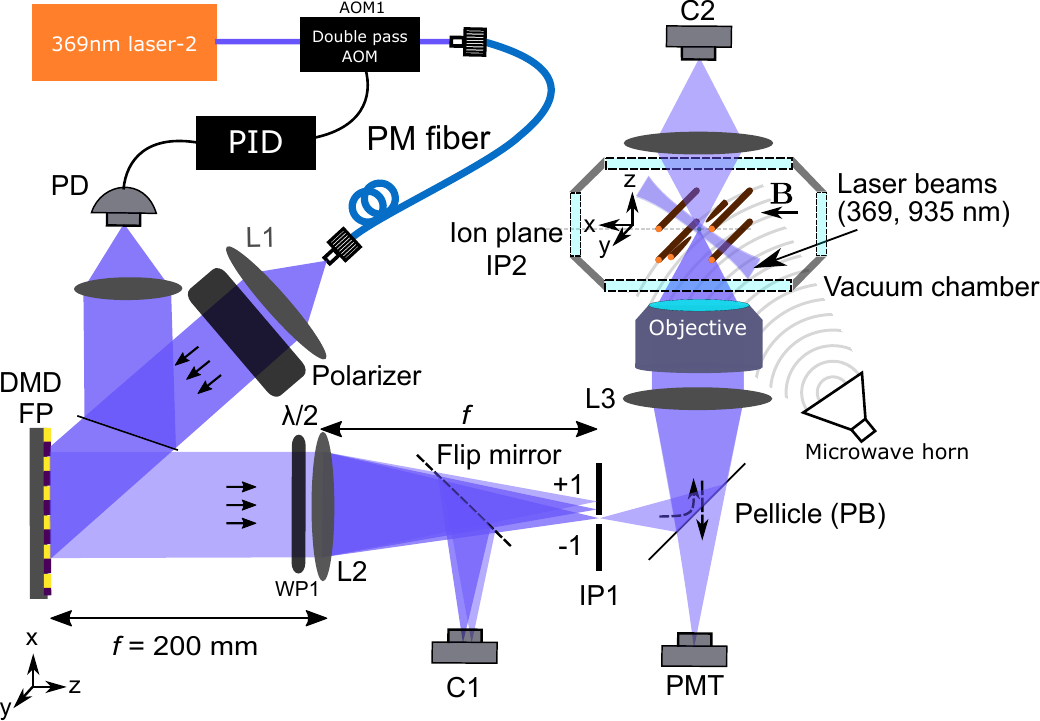}
    \caption{Experimental setup}
    \label{fig:sup_fig1}
\end{figure*}
\section{Detailed experimental setup}
Our apparatus (Fig.  \ref{fig:sup_fig1}) consists of \yb ions trapped in a four-rod paul trap with radial secular frequencies ($\omega_x$,$\omega_z$) of around $2\pi\times$ 1.1 MHz and axial trap frequencies ($\omega_y$) of $2\pi\times$ 270 kHz. 
The ground state hyperfine levels $S_{1/2}\ket{F=0,m_F=0}$ and $S_{1/2}\ket{F=1,m_F=0}$ (separated by 12.642813 GHz) are assigned as the $\ket{\downarrow}$ and $\ket{\uparrow}$ of effective spin-$1/2$ particle, respectively.
A magnetic field (B) perpendicular to the ion chain provides the quantization axis and a Zeeman splitting of $\Delta_{zm} = 2\pi\times 3.25$ MHz between $S_{1/2}\ket{F=1,m_F=0}$ and $S_{1/2}\ket{F=1,m_F=1}$ levels.
Global Doppler cooling, state-detection, and optical pumping all derived from a laser source (369nm-laser-1) along with repump beams (935 nm) are illuminated onto the ions in the XY plane. 
The fluorescence from the ions is collected through an in-house built objective onto a PMT(Hamamatsu: H10682-210) through a pellicle beam splitter(45:55) (Thorlabs: BP145B5). 
The state-dependent fluorescence transmitted after the trap is also monitored using a CMOS camera C2 (FLIR: Blackfly S BFS-PGE-04S2M) as shown in Fig.  \ref{fig:sup_fig1}.
A microwave field drives the $\ket{\downarrow}$ to $\ket{\uparrow}$ transition.
A probe beam (along z) is illuminated onto the ions through an addressing system of effective numerical aperture(NA) of 0.16(1).
This probe beam is resonant to \D0 or \D1 transitions to perform site-selective state reset or measurement.
We use another 369nm(369nm-laser-2) source for the probe beams whose frequency can be independently tuned (without affecting the global detection and cooling beams) to either \D0 or \D1 transitions. 
An acoustic-optic modulator (AOM1) in a double pass configuration, placed after the 369nm-laser-2, is used as a switch with precise timing and power control for the probe light.
The light is then coupled to a PM fiber which is then expanded using a single lens(L1) and is polarization-cleaned using a polarizer. 
The light is sampled onto a photodiode (PD) that is used to stabilize the intensity fluctuations using PID feedback to the AOM.
The polarization-cleaned and power-stabilized light from the PM fiber illuminates a Digital Micromirror device (DMD) (Visitech Luxbeam 4600 DLP) placed in the Fourier plane. 
A motorized $\lambda /2$ waveplate(WP1) is placed after the DMD to control the final polarization of the light. 
The DMD is programmed with an aberration-corrected amplitude hologram generated from an iterative Fourier transform algorithm (IFTA)(Sec. \ref{sec:hologram_generation}) to produce a Gaussian beam of waist $w$=1.50(5) $\mu m$ in the ion plane (IP2).
The negative first-order beam diffracted from the hologram on DMD is then relayed to the ion through the reflection of the pellicle. 
A flip mirror placed before the intermediate image plane IP1 is used to image the IP1 onto a camera C1 for initial characterization.
Due to the limitations of our trap parameters, such as maximum electrode voltage, we could trap two \yb ions with an inter-ion spacing no smaller than 9 $\rm{ \mu m} = 6w$.

\begin{figure}[t!]
    \centering
    \includegraphics[width =0.4 \textwidth]{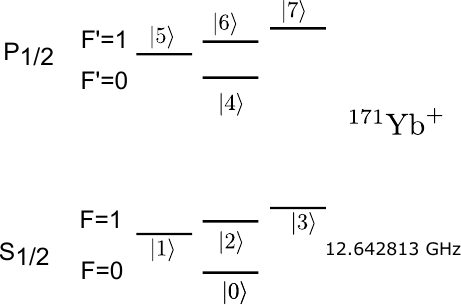}
    \caption{Encoding for $S_{1/2}$ and $P_{1/2}$ energy levels of \yb ion}
    \label{fig:sup_fig3}
\end{figure}

Despite designating the ground state hyperfine levels of \yb ions ($S_{1/2}\ket{F=0,m_F=0}$ and $S_{1/2}\ket{F=1,m_F=0}$) with $\ket{\downarrow}$ and $\ket{\uparrow}$ of an effective spin-$1/2$ particle, the measurement and reset processes involve additional states.
Consequently, the ions may ultimately occupy the states $S_{1/2}\ket{F=1,m_F=-1}$ and $S_{1/2}\ket{F=1,m_F=1}$ outside the Hilbert space of the qubit.
To model the dynamics of the ion pertaining to this work, we account for eight levels in our Hilbert space, 4 for S$_{1/2}$ and 4 for P$_{1/2}$ (Fig.  \ref{fig:sup_fig3}).
The process of state detection mixes the ion in state $\ket{2}$ with states $\ket{1}$ and $\ket{3}$, and with $\ket{0}$ when off-resonant excitation to $P_{1/2}\ket{F=1}$ states are included.
The process of state reset mixes the ion in state $\ket{2}$ with states  $\ket{0}$, $\ket{1}$ and $\ket{3}$.
The microwave field used in the Ramsey measurements (Sec. \ref{sec:ramsey_interferometry}) only couples the levels $\ket{2}$ and $\ket{0}$.

\section{Ramsey interferometry}
\label{sec:ramsey_interferometry}
To estimate the Fidelity \adjfid(main text Eq. 1), we use a set of Ramsey measurements to characterize the $\paqm$ caused by the probe beam parked at a distance $d$ from the ion-1. (Main text Fig.   2).
Each set of measurements is initialized by a sequence of Doppler cooling for 2.5 ms, optically pump(global) to $\ket{\downarrow}$ ($\ket{0}$) for 20 $\mu s$.
The probe light is illuminated for a time $T$ between two microwave $\pi/2$ pulses (detuned from a transition $\ket{\downarrow}$ to $\ket{\uparrow}$ by $\Delta_{uw}=2\pi\times10$ kHz) for a duration of about 6 $\mu s$ each.
A detection step follows where the ions are illuminated by a global detection beam for 1.5 ms, during which the state-dependent fluorescence from the ions is collected using a PMT.
Each such experiment is repeated 200 times, and the PMT counts are averaged over.
The averaged PMT counts are then normalized using measured counts from preparing $\ket{\downarrow}$ and $\ket{\uparrow}$ states.
The normalized fluorescence($\approx P(\ket{\uparrow})$) oscillates at a frequency of 10 kHz as the time $T$ is varied.
We denote the contrast of these oscillations by $R_c(T)$.
To extract the characteristic decay time($T_2^*$) of the Ramsey contrast $R_c(T)$ for a given configuration of $d$, these Ramsey measurements are done with varying T (main text Fig.   2a).
Using the preliminary coarse estimate of Ramsey contrast, $T_2^*$ is roughly estimated, and the time interval between 10\textmu s and  $2T_2^*$ is divided into five intervals, with each interval containing 21 data points in a span of 200 \textmu s.

\begin{figure*}[t!]
    \centering
    \includegraphics[width =0.9 \textwidth]{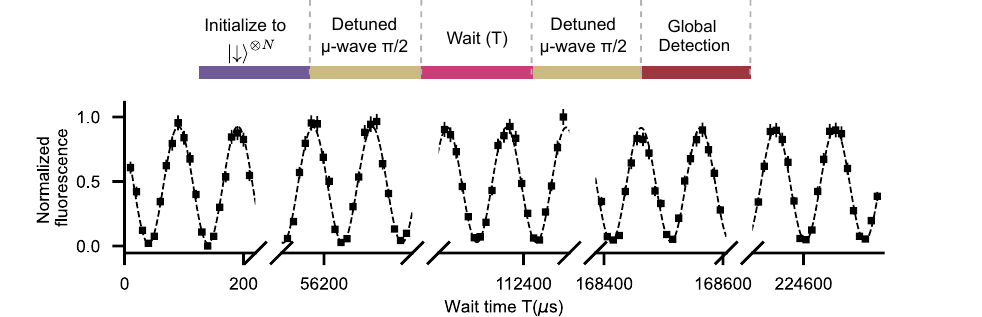}
    \caption{Ramsey measurements without probe beam}
    \label{fig:sup_fig2}
\end{figure*}

After the Ramsey measurements for these five intervals, the PMT counts are fit using the following function to extract the $T_2^*$.
\begin{equation}
\begin{aligned}
f(T)_{T_2^*,\alpha,\beta,\phi,C} =& \sin^2(\omega T+\phi)(\alpha e^{-T/T_2^*}) \\
                                 &+\beta(1-e^{-T/T_2^*})+C
\end{aligned}
\end{equation}
As a baseline measurement, we characterize the Ramsey measurements with no probe beam during the wait time (Fig.   \ref{fig:sup_fig2}) and estimate that the $T_2^*$ is much larger than 200ms.

\section{Fidelity estimation}
\begin{center}
\begin{figure}[t!]
    \centering
    \includegraphics[width =0.5 \textwidth]{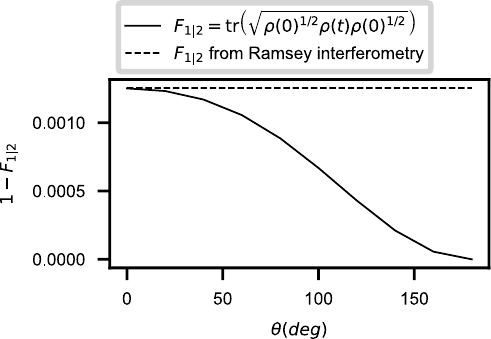}
    \caption{Infidelity( 1-\adjfid) as a function of Bloch angle $\theta$.
    Here the ion-1 is initialized in state $\rho(0) = \ket{\psi}\bra{\psi}$ where $\ket{\psi} = cos(\theta/2)\ket{2} + sin(\theta/2)\ket{0}$.
    The \adjfid(Eq. \ref{eqn:fidelity_metric}) is measured after a weak state-detection (\D0) light 
    of intensity $I=5\times10^{-5}I_{\mathrm{sat}}$ and polarization $I_{\pi}/I = \frac{1}{3}$ is applied on ion-1 for $11us$.
    For comparison, the infidelity ($1-F_{1|2}$)is shown(dotted line) from Ramsey interferometry(Sec. \ref{sec:ramsey_interferometry}) where a probe light of similar parameters as above is illuminated on ion-1 during the wait time.
    }
    \label{fig:sup_fig4}
\end{figure}
\end{center}

To quantify how well the quantum state of ion-1 is preserved after an operation on ion-2, we use the fidelity metric (\adjfid)  \cite{Nielsen2012QuantumInformation} defined as 
\begin{equation}
\label{eqn:fidelity_metric}
{F_{1|2}}(t) = \mathrm{tr}\left(\sqrt{\rho(0)^{1/2}\rho(t)\rho(0)^{1/2}}\right)
\end{equation}
where $\rho(0)$ and $\rho(t)$ denote density matrix operators of ion-1 (assuming unentangled with ion-2) before and after a state-reset or measurement operation (performed for time t) on ion-2, respectively. 
This metric yields a different value based on the initial state of ion-1, and using numerical simulations (Sec. \ref{sec:master_equations_sim}), we find that $\rho(0) = \ket{2}\bra{2}$ represents the worst case scenario (Fig.   \ref{fig:sup_fig4}).

By analytically solving the master equation of the system we find that the Ramsey fringe contrast $R_c(T)$ could be used to estimate the worst-case fidelity of ion-1 after an operation ion-2 for a time (T) using 
\begin{equation}
\label{eqn:f21_metric}
F_{1|2}(T) = \frac{2}{3}R_c(T) + \frac{1}{3}
\end{equation}
To derive the above expression analytically, we assume that the intensity of probe light decohering the ion-1 is very weak, that it causes a low probability of accidental measurement $\paqm<<1$. 
In this limit, consider the density matrix of ion-1 in a reduced Hilbert space with only ($\ket{0},\ket{1},\ket{2},\ket{3}$) states.
We model the action of probe light using the collapse operators $C_n = \sqrt{\gamma_n} \ket{i}\bra{j}$ where $i,j\in\{0,1,2,3\}$ where $\gamma_n$ is the rate of collapse.
The collapse operators and their rates depend on the transition the probe is driving and its polarization.
For example, in state detection, only the probe with $\pi$ polarization causes the AQM of ion-1. In the limit $\paqm<<1$ with $\gamma<<1$ representing the rate of $\paqm$ we use the the following collapse operators

\begin{center}
\begin{tabular}{ |c|c|c| } 
 \hline
 Transition  & Polarization & Collapse operators \\ 
 \hline
 \D0 & $\pi$ & $\sqrt{\frac{\gamma}{3}}\ket{2}\bra{2}$, $\sqrt{\frac{\gamma}{3}}\ket{1}\bra{2}$, $\sqrt{\frac{\gamma}{3}}\ket{3}\bra{2}$\\ 
 \D1 & $\sigma^+$ & $\sqrt{\frac{\gamma}{3}}\ket{2}\bra{2}$, $\sqrt{\frac{\gamma}{3}}\ket{3}\bra{2}$, $\sqrt{\frac{\gamma}{3}}\ket{0}\bra{2}$\\ 
 \D1 & $\sigma^-$ & $\sqrt{\frac{\gamma}{3}}\ket{2}\bra{2}$, $\sqrt{\frac{\gamma}{3}}\ket{1}\bra{2}$, $\sqrt{\frac{\gamma}{3}}\ket{0}\bra{2}$\\ 
 \hline
\end{tabular}
\end{center}

For ion-1 initialized in $\rho(0) = \ket{2}\bra{2}$ state, the final state of the ion-1 after the AQM due to weak probe for a time $t$ is calculated by analytically solving the Lindblad master equation
\begin{equation}
\label{eqn:lindblad_master_equation}
\begin{aligned}
\dot\rho(t)=&-\frac{i}{\hbar}[H_{atom},\rho(t)]\\
            &+\sum_n \frac{1}{2} \left[2 C_n \rho(t) C_n^\dagger - \rho(t) C_n^\dagger C_n - C_n^\dagger C_n \rho(t)\right]
\end{aligned}
\end{equation}
Here, $H_{atom}$ in interaction picture is given by $$H_{atom} = -(\Delta_{uw}+\Delta_{zm})\ketbra{1}{1} -\Delta_{uw}\ketbra{2}{2}-(\Delta_{uw}-\Delta_{zm})\ketbra{3}{3}$$
Here $\Delta_{uw},\Delta_{zm}$ denote the detuning of the microwave field and Zeeman splitting, respectively.
From the solution, we find the fidelity \adjfid to be
\begin{equation}
\label{eqn:fidelity_func_time}
F_{1|2}(t) = \sqrt{\rho_{22}(t)} = e^{-\frac{1}{3}\gamma t} \approx 1-\frac{1}{3}\gamma t
\end{equation}
Similarly, after the Ramsey experiment (Sec. \ref{sec:ramsey_interferometry}), the normalized is given as
\begin{equation}
    \rho_{22}(t) = \frac{1}{4} + \frac{1}{4}e^{-1/2\gamma t-i\Delta_{uw} t} + \frac{1}{4}e^{-1/2\gamma t+i\Delta_{uw} t} + \frac{1}{4}e^{-2/3\gamma t}
\end{equation}
The Ramsey fringe contrast is then given by $R_c(T) = \rho_{22}\left(\frac{(2m+1)\pi}{\Delta_{uw}}\right)-\rho_{22}\left(\frac{2m\pi}{\Delta_{uw}}\right)$ for a positive integer $m$ and assuming $\gamma<<\Delta_{uw}$ we get 
\begin{equation}
\label{eqn:Rc_func_time}
    R_c(T) \approx e^{-\frac{1}{2}\gamma t} \approx 1-\frac{1}{2}\gamma t    
\end{equation}
Combining Eq. \ref{eqn:fidelity_func_time} and \ref{eqn:Rc_func_time} we get \ref{eqn:f21_metric}. 
Further, the Ramsey fringe decays exponentially with a characteristic time $T_2^*$ which leads to 
\begin{equation}
\label{eqn:f21_metric_decay}
F_{1|2}(T) = \frac{2}{3}e^{-(T/T_2^*)} + \frac{1}{3}
\end{equation}

\section{Calibrations}
\subsection{FP aberration phase profile calibration} 
\label{sec:aberration_phase_profile_calibration}
\begin{figure*}[t!]
    \centering
    \includegraphics[width =1 \textwidth]{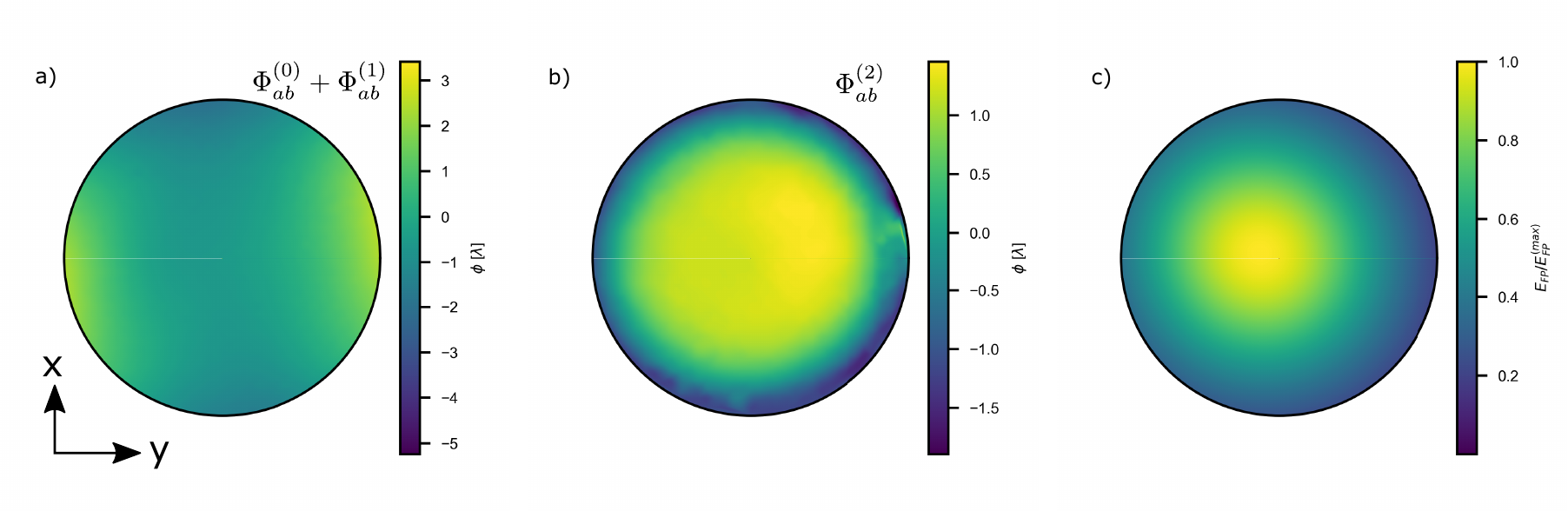}
    \caption{
    FP phase and amplitude profile. a)The aberration phase profile was measured using camera C1.
    b) The aberration phase profile is measured using the ion at IP2.
    For a-b, the piston and tilt terms are removed from the measured phase profiles, and the profile is further smoothened and interpolated.
    c) The scaled amplitude profile measured at IP2. The measured amplitude profile is smoothened, interpolated, and fit to 2D Gaussian.
    }
    \label{fig:sup_fig6}
\end{figure*}

We characterize optical aberrations in the entire beam path in terms of a Fourier plane (FP) phase map.
The optical aberrations till IP1 ($\Phi^{(0)}_{ab}+\Phi^{(1)}_{ab}$) are characterized using the camera C1 \cite{Shih2021ReprogrammableControl} as a sensor to measure the relative optical phase between two FP `patches'(Fig.   \ref{fig:sup_fig6}a).
The optical aberrations from IP1 to IP2 ($\Phi^{(0)}_{ab}$) are measured using a single ion as a sensor (see main text methods)(Fig.   \ref{fig:sup_fig6}b).
The phase profile $\Phi^{(0)}_{ab}+\Phi^{(1)}_{ab}+\Phi^{(2)}_{ab}$ is used to compensate for optical aberrations using an iterative Fourier transform algorithm (IFTA) \cite{Shih2021ReprogrammableControl} to create a diffraction-limited gaussian beam spot in IP2.

\subsection{Fourier plane intensity profile calibration} 
The incident light on DMD from L1 is nonuniform and has a Gaussian intensity profile. Further, the pellicle beam splitter (PB) has an angular dependence on reflection. The effective intensity profile on the FP is measured using an ion in IP2 as a sensor. 
The ion is prepared in $\ket{2}$ state, and the optical pumping light from DMD is used to pump to $\ket{0}$ state for a fixed time.
The value of the intensity of probe light reflected by a circular patch(30 pixels diameter) on DMD is inferred from the decrease in ion fluorescence.
This measurement is repeated for different phase-corrected (Sec. \ref{sec:aberration_phase_profile_calibration}) patches on the DMD to construct an effective intensity profile.
The intensity profile is then smoothened and interpolated, and a square root of the intensity profile is used as the amplitude profile of the incident electric field.
This amplitude profile is further used as an input to IFTA hologram generation algorithm \cite{Shih2021ReprogrammableControl}.

\begin{figure}[t!]
    \centering
    \includegraphics[width =0.4 \textwidth]{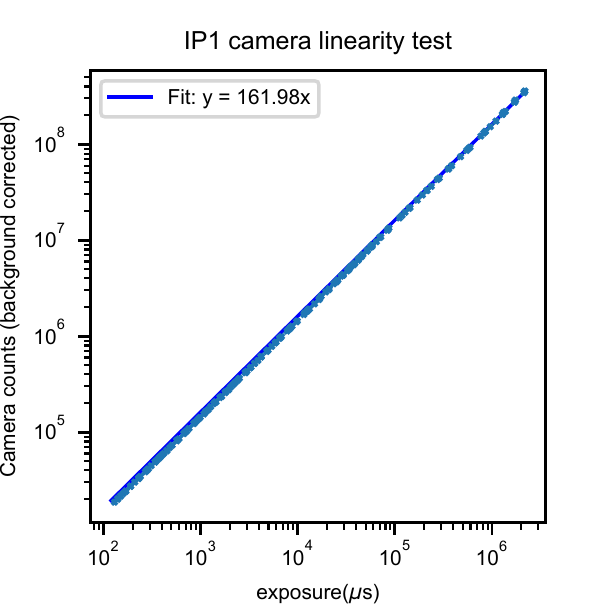}
    \caption{
    C1 camera linearity calibration
    }
    \label{fig:sup_fig7}
\end{figure}

\subsection{Relative Intensity calibration} 
The intensity of the probe light illuminating the ion through DMD is controlled by adjusting the RF power of the AOM1(Fig.   \ref{fig:sup_fig1}). 
The RF power vs. intensity of the light is calibrated using camera C1. 
To ensure accurate reporting, the linearity of C1’s exposure time is confirmed over four orders of magnitude from 100 $\mu$s to 5 s (Fig.   \ref{fig:sup_fig7}).
The linearity of the RF source power setting and the RF power output is calibrated using a spectrum analyzer.
The pellicle placed after the IP1 has a polarization-dependent transmission profile that is calibrated using camera C2 and compensated using the AOM.
This calibration also gives a relative measurement connecting the attenuated and unattenuated probe beam intensity(main text Fig.   3b and Fig.  4b) through the camera's exposure time and pixel intensity (at a fixed gain).

\subsection{Absolute intensity and polarization calibration}
The intensity of the probe beam (calibrated using C1) on the ion is calibrated with respect to the saturation intensity ($I_{\mathrm{sat}}$) of the ion. 
A series of optical pumping experiments are done with the probe using varying calibrated power and input polarization(varied using the $\lambda/2$ waveplate WP). 
These experiments are then fit using numerical simulations to extract the absolute intensity and polarization of the light illuminating the ion.

\subsection{Probe beam position and size calibration}
\label{sec:calibration_beam_position}
\begin{figure}[t!]
    \centering
    \includegraphics[width =0.5 \textwidth]{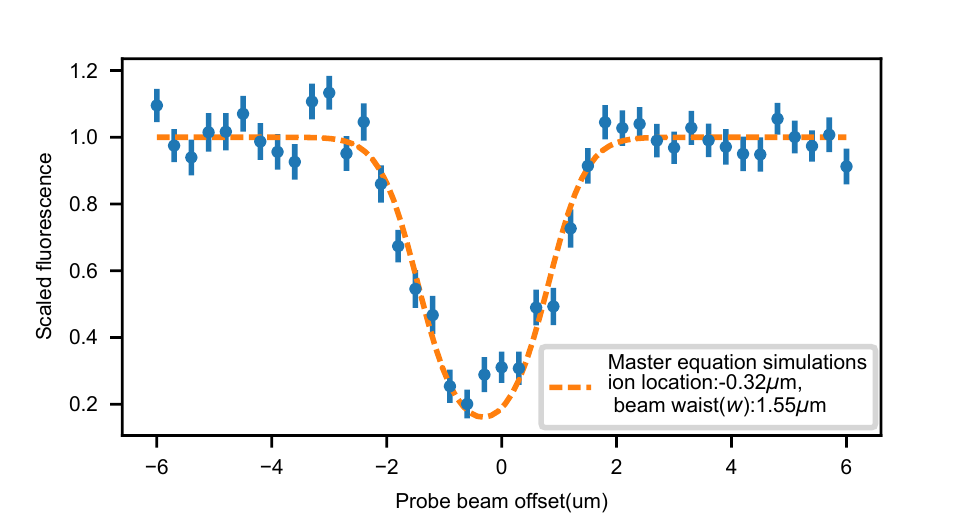}
    \caption{
    Probe beam position and size calibration
    }
    \label{fig:sup_fig5}
\end{figure}

The position of the probe beam and its beam waist is calibrated by using a single ion as a sensor for the intensity.
The ion is initialized in state $\ket{2}$, and the probe beam (state-reset) illuminates the ion for a time smaller than the optical pumping time, followed by a state measurement.
The dependence of ion fluorescence as a function of beam position is used to extract the relative beam position and the beam waist (Fig.  \ref{fig:sup_fig5}).
Here the position of the probe beam is changed by programming the hologram on the Digital micromirror device (DMD) to generate a shifted Gaussian beam.
This procedure is regularly done before every set of experiments to fix the slow drift of the relative position of the beam to the ion.
We measured that the probe beam drifts by about 0.20(15) \textmu m over the period of 15 min (a single Ramsey measurement).

\subsection{Length scale calibration}
The imaging system's effective focal length ($\approx24$mm) translates the known length scale in the Fourier plane (FP) to the length scale in IP2.
We find the relative beam positions of two ions in a trap using an experiment similar to (Sec. \ref{sec:calibration_beam_position}).
The inter-ion spacing could be calculated from the difference between the estimated relative beam positions of the two ions.
This estimated inter-ion spacing is compared to an estimation of equilibrium positions (estimated from the measured trap frequencies) to further calibrated the system's effective focal length of the system.
We could calibrate the length scale in IP2 using this method to within 5\% accuracy.

\subsection{Frequency calibration}
The relative shift of the laser frequency is calibrated by tuning the laser to the optical pumping transition and maximizing its pumping efficiency onto the ion.

\section{Setup for numerical simulations of Lindblad master equation}
\label{sec:master_equations_sim}
To model the dynamics of the ion pertaining to this work, the relevant levels are within the S$_{1/2}$ and the P$_{1/2}$ manifolds (Fig.  \ref{fig:sup_fig3})
The Hamiltonian, describing the interaction with radiation, accounts for couplings due to optical pumping, state detection, and the microwave. 
For the purpose of efficient numerical simulations, it is useful to remove the time dependence through a rotating transform ($U(t)$)  \cite{Einwohner1976AnalyticalApproximation} such that $H_{rot}=U H U^\dag - U\frac{d}{dt}U^\dag$. 
We find that when the optical pumping, detection, and microwave couplings are monochromatic, the solution of U exists, and we use this to remove the time dependence from our total Hamiltonian.
With the time dependence of the Hamiltonian accounted for, the time evolution of the density matrix can be determined by solving the Lindblad master equation(Eq. \ref{eqn:lindblad_master_equation}) with appropriate collapse operators due to the spontaneous emission. 
Using such numerical simulations, the evolution of the density matrix is calculated in a Ramsey interferometry(Sec. \ref{sec:ramsey_interferometry}).
$T_2^*$ is extracted from the simulations, and the dependence as a function of input intensity on ion-1 is calculated.
This dependence is used to extract the intensity crosstalk $I_X$ from measured $T_2^*$ of Ramsey measurements.

\subsection{Rabi fequencies}
The rabi frequencies for a simple 2-level system are set according to the formula: 

\begin{equation*}
    \frac{I}{I_{\mathrm{sat}}}=\frac{2\Omega^2}{\Gamma^{2}},
\end{equation*}

where  $I$ is the intensity of the laser, $I_{\mathrm{sat}}$ is the saturation intensity, and $\Gamma$ is the spontaneous emission rate of the transition. In our case, we are interested in finding the rabi frequency pertaining to a specific transition i.e  
\begin{equation*}
\begin{aligned}
    \Omega_{F,m_{F},F',m_{F'}  } = \frac{ \mel{ F,m_{F}}{d \cdot E}{F',m_{F'} } }{\hbar}
\end{aligned}
\end{equation*}
Applying the Wigner-Eckart theorem, we get

\begin{equation*}
\begin{split}
\mel{ F,m_{F}}{d_q}{F',m_{F'} } = \braket{ F,m_{F}}{F',m_{F'},1,q } \mel{F}{|d_q|}{F'} 
\end{split}
\end{equation*}
Now, the reduced matrix element can be further broken down as 

\begin{equation*}
    \begin{split}
         \mel{F}{|d|}{F'}  =& \mel{J I F}{|d|}{J' I F'} \\
    =& \mel{J}{|d|}{J'} (-1)^{F'+J+1+I}  \\
     & \sqrt{ (2F' + 1)  (2J + 1)}\Sixj{J}{J'}{1}{F'}{F}{I}
    \end{split} 
\end{equation*}

The reduced matrix element between the J levels can simply be calculated from the decay rate of the excited state using Fermi's golden rule as follows:

\begin{equation*}
    \Gamma_{Jg Je} = \frac{\omega_0^3}{3 \pi \epsilon_0 \hbar c^3} \frac{2J_g + 1}{2J_e + 1} |\mel{J_g}{|d|}{J_e}|^2
\end{equation*}

Since it is commonly used in \yb literature, we introduce the saturation intensity as defined for the $^2$S$_{1/2}$ to $^2$P$_{1/2}$ ignoring the internal structure:

\begin{equation*}
    I_{\mathrm{sat}} = \frac{\pi \Gamma c h}{3 \lambda^{3}}
\end{equation*}
 Combining these equations along with  $|E| = \sqrt{ I / 2 c \epsilon_0 }$ we get

 \begin{equation*}
 \begin{split}
     \Omega^2_{ F,m_{F},F',m_{F'} } &= \frac{I}{I_{\mathrm{sat}}} \frac{\Gamma_{ Jg Je}^2 }{ 2 } |\braket{ F,m_{F}}{F',m_{F',1,q} }|^2 \times \\
     & (2F' + 1)  (2J + 1)   \Sixj{J}{J'}{1}{F'}{F}{I}^2 \times 
       \frac{2J_e + 1}{2J_g + 1} 
 \end{split}
\end{equation*}

In the case of \yb, for all the allowed transitions between  $^2$S$_{1/2}$ and $^2$P$_{1/2}$, the second line of the above expression evaluates to 1/3, leaving us with a particularly simple expression for the rabi frequency 
\begin{equation}
    \Omega^2 = \frac{I}{I_{\mathrm{sat}}} \frac{\Gamma^2}{6}
\end{equation}

\section{Process qubit detection efficiency}
Consider a probe beam resonant to \D0 transition illuminates the process qubit for a time $t$. 
The scattered light from the process qubit is collected using a detector of efficiency $\varepsilon _{\mathrm{sys}}$.
The state of the qubit is inferred to be $\ket{\uparrow}$ if the detector registers a single count.
We use the approach presented in Ref. \cite{Noek2013HighQubit, MarkActon2008DetectionAtoms, Crain2019High-speedDetectors} to estimate the detection fidelity.
The error in detecting the $\ket{\uparrow}$ is given by the probability of detecting no photons when the qubit is initialized in $\ket{\uparrow}$ state \cite{Crain2019High-speedDetectors}.
\begin{equation}
\begin{aligned}
{P_{t,\ket{\uparrow}}}(n = 0) =& \frac{{R_{\mathrm{d}}}}{{\varepsilon _{{\mathrm{sys}}}R_{\mathrm{o}} + R_{\mathrm{d}}}}e^{ - R_{{\mathrm{bg}}}t}\left[ {1 - e^{ - (\varepsilon _{{\mathrm{sys}}}R_{\mathrm{o}} + R_{\mathrm{d}})t}} \right] \\
&+ e^{ - R_{\mathrm{d}}t}e^{ - ((\varepsilon _{{\mathrm{sys}}}R_{\mathrm{o}} + R_{{\mathrm{bg}}})t}
\end{aligned}
\end{equation}
Here $R_o$ denotes the scattering rate of state $\ket{\uparrow}$, $R_b$ denotes the bright pumping rate, and $R_d$ denotes the dark pumping rate \cite{Noek2013HighQubit}.
We note a plausible typo in the above expression in Ref. \cite{Crain2019High-speedDetectors} with a prefactor $\varepsilon _{\mathrm{sys}}$ missing in the exponential of the second term.
Similarly, the probability of detecting no photons when the qubit is initialized in state $\ket{\downarrow}$ \cite{Crain2019High-speedDetectors}
\begin{equation}
\begin{aligned}
P_{t,\ket{\downarrow}}(n = 0) =& \frac{{R_{\mathrm{b}}}}{{\varepsilon _{{\mathrm{sys}}}R_{\mathrm{o}} - R_{\mathrm{b}}}}e^{ - R_{{\mathrm{bg}}}t}\left[ {e^{ - R_{\mathrm{b}}t} - e^{ - \varepsilon _{{\mathrm{sys}}}R_{\mathrm{o}}t}} \right] \\
&+ e^{ - R_{\mathrm{b}}t}e^{ - R_{{\mathrm{bg}}}t}
\end{aligned}
\end{equation}
The average fidelity of state-detection of the process qubit is given by 
$$F = \frac{\left(1 - {P_{t,\ket{\uparrow}}}(n = 0)\right) + P_{t,\ket{\downarrow}}(n = 0)}{2}$$

\section{Error Analysis}
We use the bootstrapping method to get the standard error of the fitted parameters of the population transfer or the decoherence time.
We randomly resampled the dataset of the same DMD probing duration with replacement. 
The resampled dataset is used for extracting the fitted parameters. 
Repeating the resampling and fit process creates the empirical distribution of the fitting parameters.
In this work, we repeatedly resample the dataset 20 times, and the standard deviation of the 20 fitting parameters of the resampled datasets is used as the error. 
With the error of the $\tau(\mathrm{ion2})$ and $T_2^*(\mathrm{ion1})$ time through bootstrapping, the error of estimating fidelity can be derived with error propagation.

\section{Algorithm for hologram generation }
\label{sec:hologram_generation}
In this work, we used an iterative Fourier transform algorithm \cite{Shih2021ReprogrammableControl} to calculate the required hologram to display on the DMD.
However, we improved the power efficiency(see `power efficiency' of supplementary information in Ref.  \cite{Shih2021ReprogrammableControl}) by scaling up the target field with $\frac{4}{\pi}$ during the IFTA binarization process.

This new improvement is based on the fact that for a square wave $x(t)$ its fundamental mode $\sin(\omega t)$ can have a coefficient greater than $1$.
\begin{align}
    x(t) &= \frac{4}{\pi} \sum_{k=1}^{\infty} \frac{\sin ((2k-1)\omega t)}{2k-1} \\
    &= \frac{4}{\pi} \left ( \sin(\omega t) + 
    \frac{1}{3} \sin(3\omega t) + 
    \frac{1}{5} \sin(5\omega t) + ...
    \right )
\end{align}
Even though DMD can have binarized control on the grating amplitude (0~1), a higher modulation level can be achieved.
With the new improvement, the power of the signal can be enhanced by $(\frac{4}{\pi})^2 \approx 1.6$ times, which also effectively improves the signal-to-noise background ratio.

\end{appendices}

\bibliography{mendeley-references}
\clearpage

\end{document}